\documentclass[aps,prl,groupedaddress,superscriptaddress,longbibliography,reprint,11pt,onecolumn,]{revtex4-2}
\usepackage{graphics}
\usepackage{xcolor}
\usepackage[colorlinks=true,urlcolor=blue]{hyperref}
\usepackage{ulem}
 \usepackage{lineno}

\hypersetup{
	colorlinks=true, 
	breaklinks=true, 
	urlcolor= blue, 
	linkcolor= black, 
	citecolor=black, 
	pdftitle={Rapport de stage}, 
	pdfauthor={Anonyme}, 
	pdfsubject={Simulation} 
}
\usepackage{amsmath}
\usepackage{placeins}
\begin{document}
\title{Probing thermalization and dynamics of high-energy quasiparticles in a superconducting nanowire by scanning critical current microscopy}

\author{T. Jalabert}
\email[]{thomas.jalabert@grenoble-inp.fr}
\affiliation{Univ. Grenoble Alpes, CEA, Grenoble INP, IRIG, PHELIQS, 38000 Grenoble, France}

\author{E. F .C. Driessen}
\affiliation{Institut de RadioAstronomie Millim\'etrique (IRAM), 38400 Saint Martin d'H\`eres, France}

\author{F. Gustavo}
\affiliation{Univ. Grenoble Alpes, CEA, Grenoble INP, IRIG, PHELIQS, 38000 Grenoble, France}

\author{J.L. Thomassin}
\affiliation{Univ. Grenoble Alpes, CEA, Grenoble INP, IRIG, PHELIQS, 38000 Grenoble, France}

\author{F. Levy-Bertrand}
\affiliation{Univ. Grenoble Alpes, CNRS, Grenoble INP, Institut N\'eel, 38000 Grenoble, France}

\author{C. Chapelier}
\email[]{claude.chapelier@cea.fr}
\affiliation{Univ. Grenoble Alpes, CEA, Grenoble INP, IRIG, PHELIQS, 38000 Grenoble, France}

\begin{abstract}
	
\textbf{Besides its fundamental interest, understanding the dynamics of pair breaking in superconducting nanostructures is a central issue to optimize the performances of superconducting devices such as qubits or photon detectors. However, despite substantial research efforts, these dynamics are still not well understood as this requires experiments in which quasiparticles are injected in a controlled fashion. Until now, such experiments have employed solid-state tunnel junctions with a fixed tunnel barrier. Here we use instead a  cryogenic scanning tunnelling microscope to tune independently the energy and the rate of quasiparticle injection through, respectively, the bias voltage and the tunnelling current. For high energy quasiparticles, we observe the reduction of the critical current of a nanowire and show it is mainly controlled by the injected power and, marginally, by the injection rate. Our results prove a thermal mechanism for the reduction of the critical current and unveil the rapid dynamics of the generated hot spot.}

\end{abstract}


\maketitle

The performance of superconducting devices is often limited or governed by  quasiparticle dynamics. Whereas excess quasiparticles are detrimental for superconducting microcoolers~\cite{nahum_Electronic_1994,giazotto_Opportunities_2006}, current sources in metrology~\cite{pekola_single-electron_2013}, high kinetic inductances~\cite{grunhaupt_loss_2018}, and superconducting qubits~\cite{martinis_Energy_2009,riste_Millisecond_2013,wang_measurement_2014,pop_coherent_2014,janvier_coherent_2015,hays_direct_2018,serniak_hot_2018,vepsalainen_impact_2020,karzig_quasiparticle_2021}, the generation of quasiparticles is a prerequisite for the operation of photon detectors~\cite{peacock_single_1996,goltsman_Picosecond_2001,day_broadband_2003,walsh_josephson_2021,natarajan_Superconducting_2012}. A precise knowledge of the physical mechanisms underlying the quasiparticle dynamics is thus needed for optimization of device performance. However, despite  intense research efforts~\cite{kozorezov_quasiparticle-phonon_2000,bulaevskii_vortex-assisted_2012,zotova_photon_2012,zotova_intrinsic_2014,renema_Experimental_2014,engel_Detection_2015a,vodolazov_single-photon_2017,nicolich_universal_2019,kubo_superfluid_2020,semenov_superconducting_2021}, the processes at stake in the energy relaxation of quasiparticles in current-carrying superconductors are still not well understood. One striking example is the recent proposal of an all-metal Josephson field-effect transistor (JFET) that relies on the modulation of its critical current under the application of a gate voltage~\cite{desimoni_Metallic_2018,paolucci_UltraEfficient_2018}. This proposal has given rise to considerable controversy, suggesting instead a heating effect through the injection of high-energy quasiparticles~\cite{alegria_high-energy_2021,ritter_superconducting_2021,golokolenov_origin_2021,catto_microwave_2022}. All these experiments have employed solid-state tunnel junctions with a fixed resistance which do not easily allow to disentangle current and voltage effects. We overcome these limitations with a new experimental set-up based on a very low temperature scanning tunnelling microscope (STM) working at 50 mK which allows to locally inject quasiparticles into a superconducting nanowire with a nanometer spatial resolution while simultaneously measuring its critical current. With this so-called scanning critical current microscope, we can vary independently the tunnelling current It and the energy of the injected quasiparticles eVb, where e is the electronic charge and Vb the bias voltage of the tunnelling junction. We show that for injection at bias voltages larger than the superconducting gap, the reduction of the critical current is scaling with the injected power. This scaling and the spatial study of the quasiparticles injection on the critical current are explained by a thermal mechanism. Moreover, we observe a marginal influence of the injection rate on the critical current that we interpret as a signature of the quasiparticle energy relaxation during the first tens of picoseconds after their injection.

\begin{figure}
	\resizebox{\linewidth}{!}{\includegraphics{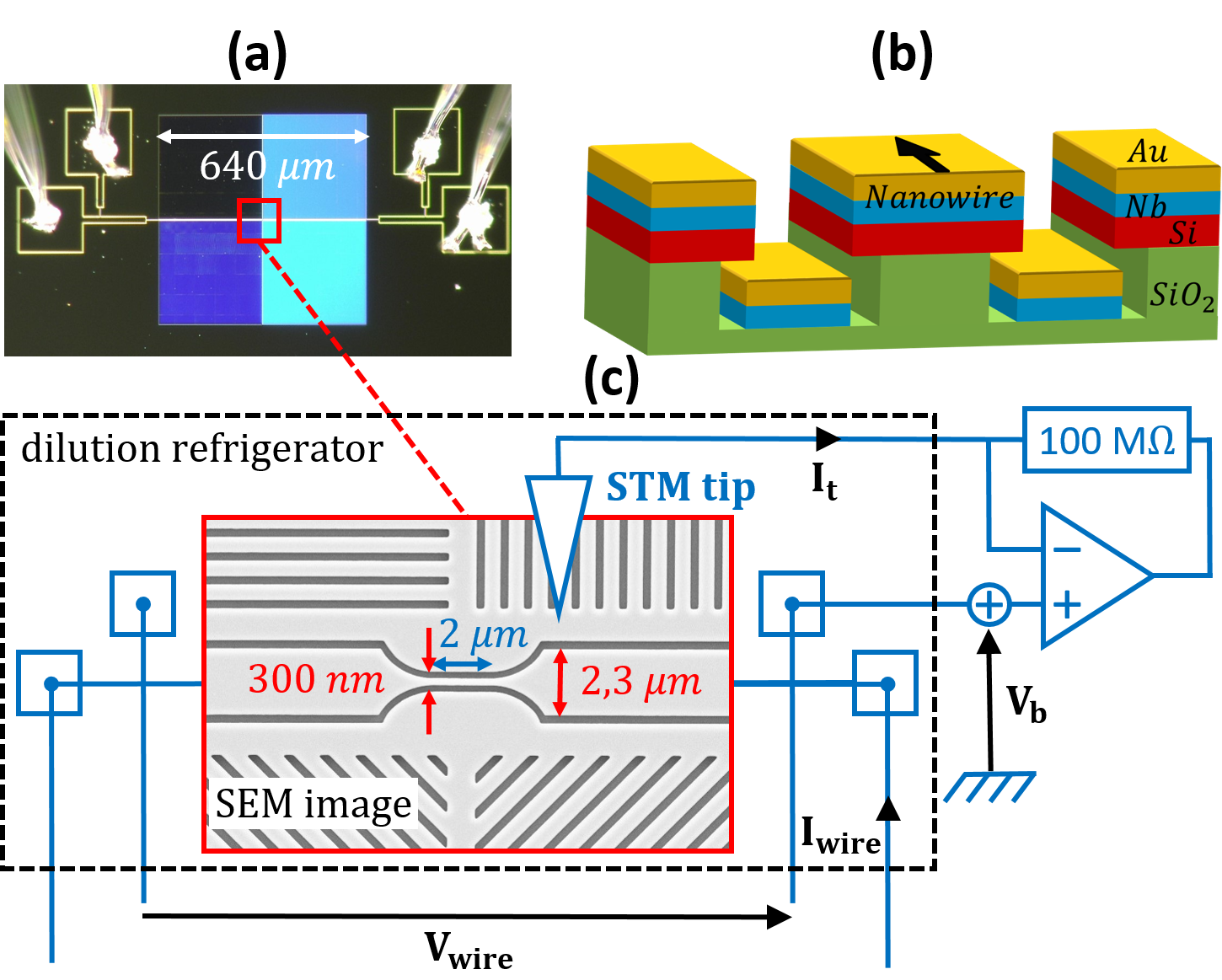}}
	\caption{\textbf{Experimental set-up} \textit{(a)} Picture of a nanowire at the center of a target of lines (blue squares) with four microbondings (at the edges). \textit{(b)} Schematic of the sectional view of a nanowire. \textit{(c)} The nanowire is scanned by a STM tip at a bias potential $V_b$ and tunnelling current $I_t$ while the $I_{wire}-V_{wire}$ characteristic  is monitored. }
	\label{setup}
\end{figure}

Contacting and locating an individual nanostructure with an STM is very challenging due to the intrinsic incompatibility of an STM with the insulating regions necessary to define a galvanically isolated nanostructure. Other research efforts have focused on combining AFM and STM~\cite{quaglio_subkelvin_2012,le_sueur_phase_2008,coissard_imaging_2022}, which intrinsically inherits the difficulties of both technologies. We take another approach in which we uniquely use our STM for both locating and measuring our nanodevice. Figure 1 presents the Scanning Critical Current Microscopy set-up. The nanowires are made of bi-layers of Nb/Au evaporated on a pre-etched Si/SiO$_2$ substrate (see Figure~\ref{setup}b). The role of the Au capping layer is to prevent oxidation of the Nb and to warrant a good tunnelling junction with the STM Pt-Ir tip. The topology of the pre-etched substrate provides a fully metallic surface compatible with the STM technique while preserving a galvanic isolation of the nanowire with the surrounding grounded film. Six superconducting nanowires with different Nb/Au nominal total thicknesses ranging between 6 nm and 10 nm were studied (see Supplementary for more details). The 300~nm~$\times$~2~$\mu$m nanowire is localized using a target formed of lithographically defined lines (see Figure~\ref{setup}c). Four electrical micro-bonded contacts are used for the transport measurements. They are connected to the nanowire through Nb/Au leads of about 320~$\mu$m long and 2.3~$\mu$m wide (see Figure~\ref{setup}a). $I_c$ is determined by ramping the current $I_{wire}$ in the nanowire while recording the voltage $V_{wire}$. The critical current is reached when an abrupt jump of  $V_{wire}$ is observed in the $I_{wire}$ - $V_{wire}$ characteristic (see Supplementary).

\begin{figure}
	\resizebox{12cm}{!}{\includegraphics{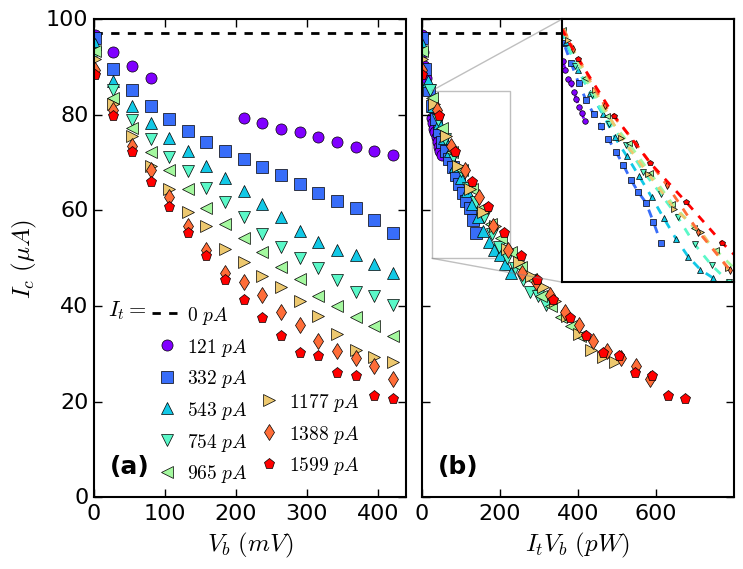}}
	\caption{\textbf{Critical current as a function of bias voltage and injected power.} Sample N03.\textit{(a)} Critical current as a function of bias voltage $V_b$ for different tunnelling currents $I_t$ at $T=250~mK$. The dashed line indicates the critical current value $I_{c}=96.3\ \mu A$  when no quasiparticles are injected. \textit{(b)} Same data as function of $I_tV_b$. The inset is a zoom of the data in the grey rectangle}
	\label{IcvsVb}
\end{figure}

Figure~\ref{IcvsVb} presents the critical current measured when quasiparticles are injected in the middle of the nanowire. Panel~(a) displays the critical current as a function of the bias voltage $V_b$ for different tunnelling currents $I_t$. Increasing either $I_t$ or $V_b$ reduces the critical current from its maximum value $I_{c}=96.3~\mu A$ measured when the STM tip is withdrawn and no quasiparticles are injected. It is striking that the critical current is reduced by tunnelling current injections that are six orders of magnitude smaller. In Panel~(b) the same data are plotted as a function of the product $I_tV_b$ which is approximately twice the injected power $P_0$ for large $eV_b$ compared to the superconducting gap $\Delta_0=370~\mu$eV as measured by tunnelling spectroscopy. All the experimental points almost merge on a unique curve. This demonstrates that the critical current is mainly controlled by the injected power. In a thermodynamical framework, each injected quasiparticle relaxes its energy to phonons, which in turn break hundreds of Cooper pairs generating many out-of-equilibrium quasiparticles, the so-called hot spot that thermalizes through inelastic electron-phonon and electron-electron interactions. This down-conversion cascade occurs on a very short time of the order of a few picoseconds~\cite{kozorezov_quasiparticle-phonon_2000,vodolazov_single-photon_2017}. A quasi-equilibrium stationary state is reached when the injected power is balanced by the power evacuated through electronic thermal conduction and electron-phonon coupling. Such a state is characterized by a local increase of the electronic temperature below the tip, resulting in a reduced local critical current.

\begin{figure}
	\resizebox{\linewidth}{!}{\includegraphics{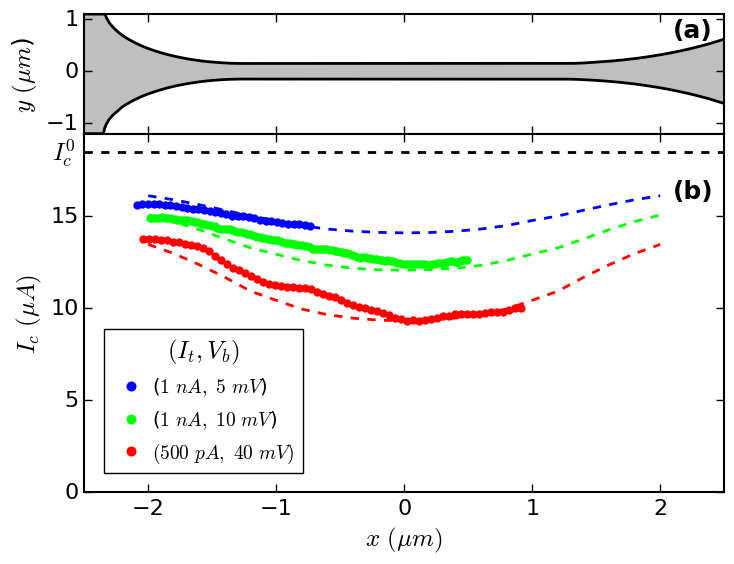}}
	\caption{\textbf{Scanning critical current microscopy.} \textit{(a)}~Nanowire shape. \textit{(b)} Dots: critical current as a function of the STM tip position $x$ along the nanowire and averaged along $y$ for different tunnelling conditions at $T=180\ mK$. Dashed lines: critical current obtained from the numerical solutions of Eq.~\ref{eq:Te} for $\Sigma=6\times10^9$ W.K$^{-5}$.m$^{-3}$. Without injection current, $I_c=18.5\ \mu A$. Sample N06.}
	\label{spatialdependance}
\end{figure}

Scanning critical current microscopy allows to go a step further by mapping the critical current $I_c(x,y)$ as a function of the STM tip position for fixed tunnelling conditions. $I_c$ is smoothly varying along the nanowire length $x$ with very little fluctuations in the transverse direction $y$ (see Supplementary). We can therefore plot $I_c(x)$ as shown on Figure~\ref{spatialdependance} where panel~(a) sketches the nanowire shape and panel~(b) presents the critical current evolution for three different injected powers. As expected, it shows that the critical current decreases further as the position of the tip moves away from the leads or as the injected power increases.

\begin{figure}
 	\resizebox{12cm}{!}{\includegraphics{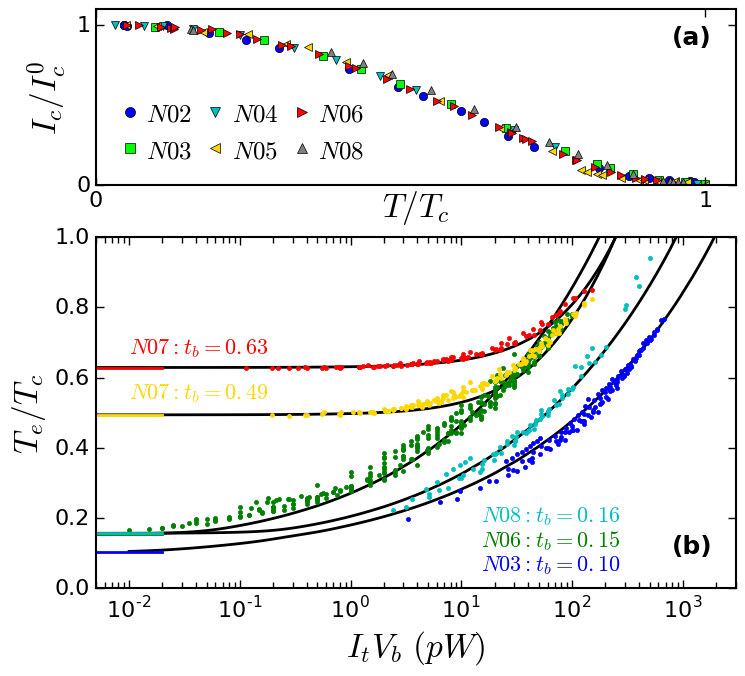}}
 	\caption{\textbf{Effect of the power injected by quasiparticles on the electronic temperature} \textit{(a)} Normalized critical current measured without quasiparticle injection $\frac{I_c(T/T_c)}{I_c^0}$ where $I_c^0$ is the critical current value at T=0 as a function of the reduced temperature. \textit{(b)} Reduced electronic temperature as a function of the quasiparticle power injected by the STM tip. For the sake of completeness in this set of experiments we varied and measured $T_b$ with the help of a heater and a thermometer glued to the sample holder. On the left of the plot, horizontal colored lines point out the corresponding reduced bath temperature $T_b/T_c$. Solid lines correspond to theoretical predictions of Eq.~\ref{eq:Te} with $\Sigma=6,6,0.8$ and $8\times10^9$~W.K$^{-5}$.m$^{-3}$ for samples N03, N06, N08 and N07, respectively.}
 	\label{figfordiscussion}
 \end{figure}

 In order to have a quantitative understanding of our data, the local electronic temperature can be estimated from the measured critical current. Without quasiparticle injection, the thermal dependence of the critical current is a universal curve as theoretically expected~\cite{kupriyanov_temperature_1980} and illustrated in Figure~\ref{figfordiscussion}(a) where measurements performed on all our nanowires are shown in reduced units. From this curve, a local electronic temperature $T_{e}$ below the tip can be associated to the measured critical current $I_c(I_t,V_b)$. It is then possible to plot the normalized electronic temperature $T_{e}/T_c$ as a function of the injected power as shown for different samples and bath temperatures in Figure~\ref{figfordiscussion}(b). At low injected power, the electronic temperature is equal to the bath temperature. At high injected power, it scales with a power law.

A minimal stationary thermal model that accounts for both the spatial and the power dependences of $T_e$ with quasiparticle injection is given by a one-dimensional heat equation: 
\begin{eqnarray}
	\nabla\big( dwk_e\frac{\partial T_{e}(x)}{\partial x}\big)=\Sigma w d(T_e^5(x)-T_{ph}^5)-P_0\delta(x-x_0)
	\label{eq:Te}
\end{eqnarray}
where  $d$ and $w$  are, respectively, the thickness and the width of the nanowire, $k_e$ is the electronic thermal conductivity, $\Sigma$ is a material-dependent electron-phonon coupling parameter, $T_{e}(x)$ is the local electron temperature and $T_{ph}$ is the phonon temperature.  We assume that the substrate acts as a phonon reservoir such that $T_{ph} = T_b$ everywhere. We have experimentally determined the limits of validity of this hypothesis (see  Supplementary). $P_0$ is the power injected by the tip at the position $x_0$. It is approximately equal to $P_0 \sim I_tV_b/2$ when the bias voltage is large compared to the superconducting gap. The left hand term is the thermal gradient along the nanowire. The electronic thermal conductivity is computed with Usadel equations~\cite{usadel_Generalized_1970} with no adjustable parameter. Its value at the critical temperature is set by the measured normal state resistivity and the Wiedemann-Franz law (see Supplementary for more details). The right hand term describes the heat flow. Its last term accounts for the heating  due to the injection of quasiparticles. The first right hand term is a cooling flow due to electron-phonon coupling in metals~\cite{wellstood_HotElectron_1994}. This power law is not strictly valid at low temperatures in superconductors. Indeed, the opening of a gap in the electronic density of states exponentially reduces the electron-phonon coupling which cannot be described by an analytical expression~\cite{guruswamy_nonequilibrium_nodate}. However, the cooling flow term is only relevant when the electronic temperature is large compared to the phonon bath temperature, i.e when the nanowire is close to its normal state. 

Solutions of the one-dimensional heat Equation~\ref{eq:Te} are compared to the spatial variation of the critical current in Figure~\ref{spatialdependance} and to the power dependence of the electronic temperature below the tip in Figure~\ref{figfordiscussion}, see dashed and solid lines respectively. The only fitting parameter is the material-dependent electron-phonon parameter $\Sigma$ that we found to be in the range of $10^9$~W.K$^{-5}$.m$^{-3}$ in agreement with previous measurements in disordered metals~\cite{wellstood_HotElectron_1994}.  The same value of $\Sigma$ accounts for both the spatial variation and the power dependence of sample N06. The remarkable agreement between the  one-dimensional heat model and all the experimental data stongly supports the thermally-driven reduction of the critical current by hot quasiparticle injection.

However, a closer look at the inset of Figure~\ref{IcvsVb}(b) shows that, although the critical current is mainly given by the injected power, there is a small systematic variation with the tunnelling current:  $I_c(P)$ fans out along different curves depending on the tunnelling current, i.e. the injection rate $\tau_{inj} = I_t/e$ of the quasiparticles. As we will show below, this is a signature of the dynamics of the injected quasiparticles and it reveals the time dependent formation of the hot spot at its early stage.
	
Following Semenov and co-workers~\cite{semenov_spectral_2005} an excess of quasiparticles $\delta N$ in  the hot spot reduces the density of Cooper pairs available to carry the superfluid current. This translates into an increase of their speed in order to sustain a constant bias current $I_{wire}$. During the down-conversion cascade, more and more quasiparticles are created so that, at one point, the Cooper pairs reach their critical velocity and the nanowire switches in the normal state. The critical current is then determined by:  
	\begin{eqnarray}
	I_c = I_c^{stat} \, (1 - \frac{\delta N}{N_0 \, \Delta_0 \, \xi \, w \,  d})
	\label{eq:Ic_vs_deltaN}
	\end{eqnarray}
where $N_0$ is the electronic density of states at the Fermi level and $\xi$ the superconducting coherence length. $I_c^{stat}$ is the stationary critical current in the limit of a continuously delivered power to the nanowire (i.e. $I_t \rightarrow \infty$).

The hot spot dynamics is governed by the balance between the proliferation of quasiparticles during the down-conversion cascade and their escape through diffusion outside the critical volume where the nanowire switches to the normal state. The former process can be accurately described by an exponential growth with a relaxation time scale $\tau_{rel}$~\cite{semenov_quantum_2001} while the critical volume of the latter is modelized by a slab of width $\xi$ across the nanowire~\cite{semenov_spectral_2005}. This leads to the time evolution : 
	\begin{eqnarray}
	\delta N = K \, \frac{1-e^{\frac{-t}{\tau_{rel} }}}{\sqrt{\pi \, D \, t}} \, \xi
	\label{eq:deltaN_evolution}
\end{eqnarray}
Here $D$ is the diffusion coefficient which is independently determined by transport measurements (see  Supplementary) and $K$ is the maximum number of quasiparticles created in the relaxation process. If one neglects phonon escape, it can be estimated as $K \simeq \frac{e V}{\Delta_0}$ where $eV$ is the energy of the tunnelling electron ($V \in \left[0,V_b\right]$). This approximation is justified since the time required for phonon to escape is much larger than the relaxation time because of phonon-trapping~\cite{kaplan_acoustic_1979}.

\begin{figure}
	\resizebox{\linewidth}{!}{\includegraphics{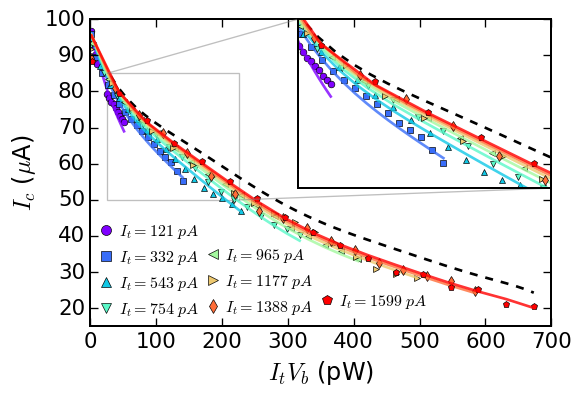}}
	\caption{\textbf{Relaxation dynamics of the injected quasiparticles.} The critical current $I_c$ at a given power $I_tV_b$ is all the more reduced that the quasiparticle injection rate is low.  Points are the same data as in Figure~\ref{IcvsVb} and solid lines are numerical fits with $\tau_{rel} = 40 ps$. The dashed line is the stationary critical current $I_c^{stat}$. The inset is a zoom at low power.}
	\label{dynamics}
\end{figure}

With this model in hand, we simulated the experiment by computing a series of N tunnelling events of electrons with an energy $eV_i$ occurring at times $t_i$. $V_i$ is uniformly distributed between $0$ and $V_b$ and the tunnelling times $t_i$ obey a Poisson distribution where on average one electron is tunnelling every $\tau_{inj}=\frac{e}{I_t}$. N has been chosen large enough so the total time of each series is in the ms range, thus comparable with the duration of the experimental step increment of the ramped $I_{wire} (t)$. Then, a single measurement of $I_c$ corresponds to the minimum of the critical current of the computed series. It appears that this minimum always occur when several tunnelling events happen in a short period of time (see Supplementary). For each tunnelling conditions of $I_t$ and $V_b$ hundreds of series have been computed in order to determine an averaged critical current and adjust $\tau_{rel}$ to fit with the data of Figure~\ref{IcvsVb}(b). Our results are shown in Figure~\ref{dynamics}. The dashed line is the stationary critical current $I_c^{stat}$ which is self-consistently determined by the fitting procedure. The fanning out of the critical current for various injection rates is due to the dynamics of the hot spot formation with a time scale of $\tau_{rel} = 40  \pm 10 ps$. It is longer than the $7 ps$ in pure Nb computed by Kozorezov and co-workers~\cite{kozorezov_quasiparticle-phonon_2000}. This difference could be attributed to a weakening of the electron-phonon scattering rate due to disorder and to the gold overlayer.

It is important to notice that in the model described above, only the growing number of out-of-equilibrium quasiparticles matters. After the time $\tau_{rel}$, they may not be fully thermalized yet with a well established Fermi-Dirac distribution at the local electronic temperature $T_e$. Nevertheless, it is interesting to notice that $\tau_{rel}$, the time of the hot spot formation, is of the same order of magnitude as $\tau_w = \frac{w^2}{4D} \simeq 40 ps$ the time required for the quasiparticles to diffuse across the width of the nanowire. This suggests that the nanowire switches to the normal state over its entire width, rather than that the normal state is triggered by current crowding around a normal-state core~\cite{renema_Experimental_2014,engel_Detection_2015a,vodolazov_single-photon_2017,semenov_superconducting_2021}. However, to be fully conclusive, the analysis should be done by solving the coupled kinetic equations for interacting quasiparticles and phonons~\cite{chang_nonequilibrium_1978} which is beyond the scope of our work.

In conclusion, we have developed a powerful new tool for studying local quasiparticle dynamics in superconducting nanostructures, in which we can independently tune tunnelling rate and quasiparticle energy. We have used this tool to show that the critical current of a nanowire is significantly reduced by a quasiparticle injection current several orders of magnitude lower. We show that this reduction is mainly due to thermal heating by quasiparticles, but that the details can only be understood by taking into account the rapid relaxation dynamics after each tunnelling of individual electron. These results have an immediate impact on the understanding of superconducting nanodevices such as JFETs and photon detectors. In addition, because of its versatility, our set-up can help designing future superconducting quantum circuit in order to strengthen them against quasiparticle poisoning.

\textbf{Methods} The substrate is made of Silicon On Insulator (SOI), with 250 nm of Si over $2 \, \mu m$  of SiO$_2$. A 360 nm layer of resist (ZEP520A) is deposited by spin coating. Then, the pattern is realized by e-beam lithography in a JEOL6300FS. Firstly, Reactive Ion Etching is carried out, resulting in an efficient anisotropic etching of the Si. Then the resist is removed by O$_2$ plasma cleaning before a chemical etching of the sample with HF vapor. The HF etching creates an undercut on the sidewall which garantees the electrical insulation of the nanowire after  evaporation of the Nb and Au thin films in an Plassys electron gun evaporator (MEB550).

\textbf{Acknowledgments} We thank M. Aprili, T. Cren, M. Houzet, T. Klein, J. Meyer and S. Sankar for fruitful discussions and G. Vinel for his help in the software development for the numerical simulation. We acknowledge funding from the CEA Eurotalents program and from the French National Research Agency  (grant ANR-16-CE30-0019-ELODIS2). This work has been partially supported by the LabEx FOCUS ANR-11-LABX-0013 and the EU\textsc{\char13}s Horizon 2020 research and innovation program under Grant Agreement No. 800923 (SUPERTED).

\textbf{Competing interests} The authors declare no competing financial interests.\\

\textbf{Author Contributions} JLT and FG prepared the samples in the clean room facility. EFCD and CC pioneered the experiment. TJ and CC performed the measurements reported in this article. TJ performed the simulations. TJ, EFCD, FL and CC analysed the data and wrote the manuscript.

\bibliographystyle{mystyle_with_doi_ter}
\bibliography{HighBias_ArXiv}

\newpage

\renewcommand\thefigure{S\arabic{figure}}
\renewcommand\thetable{S\arabic{table}}
\renewcommand\theequation{S\arabic{equation}}
\section{Supplementary Note 1: Equilibrium properties of the superconducting samples}

The samples presented in the main text are made of a thin layer of niobium capped with gold to prevent surface oxidation. Their main equilibrium properties are summarized in \autoref{table_param}.

\begin{table}[h]
	\centering
	\begin{tabular}{lccccccccc}
		\hline \hline
		Sample & Nb & Au & $T_c$ & $I_c^0$ & $\rho_N$ & $D$ & $\Delta_0$ & $\Delta_0/(k_BT_c)$ \\
		& $(nm)$ & $(nm)$ & $(K)$ & $(\mu A)$ & $(\mu \Omega.cm)$   & $(cm^2 s^{-1})$ & $(\mu eV)$ & \\
		\hline
		N03 & 5 & 5 & 2.43 & 96.3 & 18.4  & 6.8 & 370 & 1.77 \\
		N04 & 5 & 5 & 2.54 & 103 & 18.4  & 6.9 & 380 & 1.74 \\
		N05 & 3 & 3 & 1.20 & 15.7 & 24.8  & 5.1 & 220 & 2.13 \\
		N06 & 3 & 3 & 1.39 & 18.5 & 24.7  & 5.1 & 235 & 1.96 \\
		N07 & 3 & 3 & 1.72 & - & 31.3  & 4.0 & 290 & 1.95 \\
		N08 & 4 & 4 & 2.14 & 57.0 & 22.7  & 5.6 & 355 & 1.92 \\
		\hline \hline
	\end{tabular}
	\caption{Parameters of the samples: Nb~(nm) and Au~(nm) are the nominal niobium and gold thickness, $T_c$ is the critical temperature,  $I_c^0$ is the zero temperature critical current, $\rho_N$ is the normal state resistivity, $D$ is the diffusion coefficient and $\Delta_0$ the superconducting gap measured by tunneling spectroscopy.}
	\label{table_param}
\end{table}

\begin{figure}[h!]
	\centering
	\resizebox{0.65\linewidth}{!}{\includegraphics{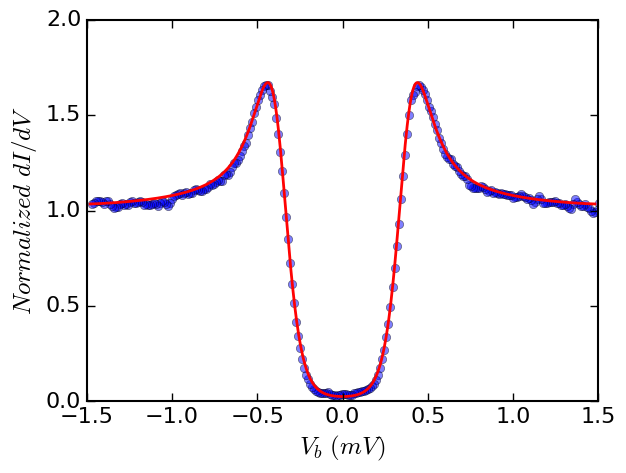}}
	\caption{\textbf{Determination of the superconducting gap by tunneling spectroscopy.} Dots: Differential conductance as a function of bias voltage $V_b$ normalized to its large bias value at $T=100$~mK. Red solid line corresponds to BCS fit (see text) with $T_{eff}=556$~mK, $\Delta=370$~$\mu$eV and $\Gamma_{in}=0.02\Delta$. Sample N03. \label{fig:delta0}}
\end{figure}

The tunneling conductance between the metallic STM tip and the sample is proportional to the local density of states~\cite{tinkham_Introduction_2004}:
\begin{equation}
	\frac{dI}{dV}(V_b)\propto\int_{-\infty}^{\infty}dE~N_s(E)\frac{\partial f(E+eV_b)}{\partial(eV_b)}
	\label{eq:didv}
\end{equation}
where $N_s(E)$ is the density of states in the sample and $f$ the Fermi-Dirac distribution function at temperature $T_{eff}$.
A typical spectrum is shown on \autoref{fig:delta0}. It displays a superconducting spectroscopy measurement that can be fitted with BCS theory~\cite{bardeen_Theory_1957} in order to extract the low temperature superconducting gap.
We use the following formula for the superconducting density of states~\cite{dynes_Direct_1978}:
\begin{equation}
	N_s(\epsilon)=
	\left\lbrace
	\begin{array}{cc}
		N_0~\left|\textnormal{Re}\left[\frac{\epsilon+i\Gamma_{in}}{\sqrt{(\epsilon+i\Gamma_{in})^2-\Delta^2}}\right]\right| & \textnormal{if } |\epsilon|>\Delta\\
		0 & \textnormal{if } |\epsilon|<\Delta
	\end{array}
	\right.
	\label{eq:DOS_dynes}
\end{equation}
where $N_0$ is the density of states at Fermi level, $\Delta$ the superconducting gap and $\Gamma_{in}$ the inelastic scattering term introduced by Dynes~\cite{dynes_Direct_1978} to account for the finite lifetime of quasiparticle excitations.

\begin{figure}[h!]
	\centering
	\resizebox{0.4\linewidth}{!}{\includegraphics{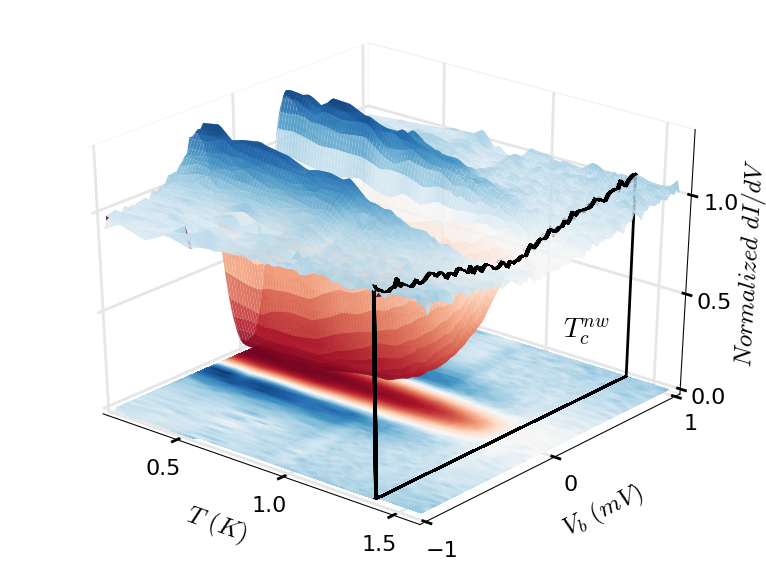}}
	\resizebox{0.4\linewidth}{!}{\includegraphics{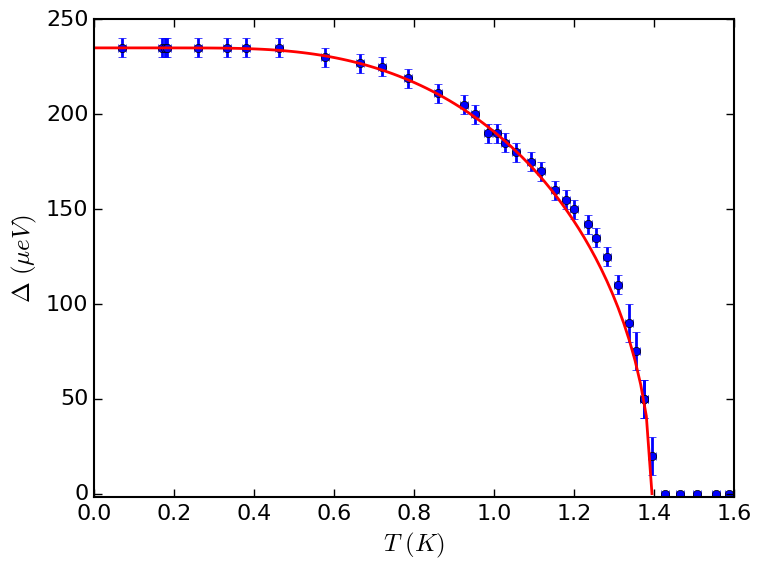}}
	\caption{\textbf{Temperature dependence of the density of states.} Sample N06. Left : normalized differential conductance as a function of bias voltage $V_b$ for different temperatures. Each spectrum is normalized to its large bias value. Black solid line corresponds to the critical temperature $T_c=1.39$~K. Right : thermal evolution of the superconducting gap. Red solid line corresponds to a BCS fit. \label{fig:DOS_T}}
\end{figure}

Because our samples are capped with a thin layer of gold, the STM measurements probe the superconducting properties induced by proximity effect in the gold layer by the niobium layer underneath. The $\Delta_0/k_BT_c$ ratio obtained from our measurements on the gold-niobium bilayer is close to the one of niobium ($\Delta_0/k_BT_c=1.83$ in niobium~\cite{novotny_Single_1975}). This indicates a strong proximity effect between gold and niobium~\cite{fominov_Superconductive_2001} which can therefore be considered as a single superconducting layer. Moreover, \autoref{fig:DOS_T} shows that the superconducting gap follows a BCS-evolution with temperature and vanishes at the critical temperature.

\newpage
\section{Supplementary Note 2: Monitoring the Current-Voltage characteristics under quasiparticle injection}

\begin{figure}[h!]
	\centering
	\resizebox{0.7\linewidth}{!}{\includegraphics{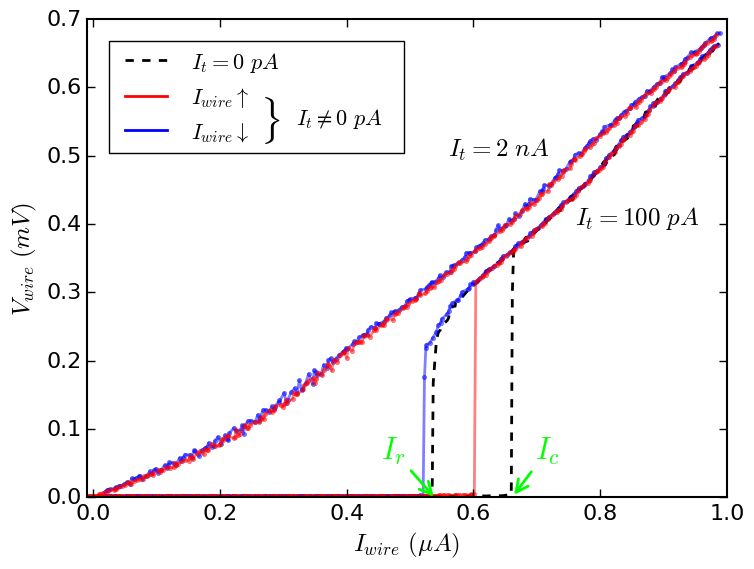}}
	\caption{\textbf{Current-voltage characteristics for different injection currents.} The wire current is raised and decreased in order to show the hysteretic loop and determine the critical current $I_c$ and the retrapping current~$I_r$. Dashed line corresponds to the situation where the STM tip is withdrawn (no tunneling current), and solid lines to $I_t=100$~pA, 500~pA and 2000~pA from right to left. All curves are recorded for a constant temperature $T=1.60$~K and bias voltage $V_b=200$~mV. Sample N07. \label{fig:I_V} }
\end{figure}

\autoref{fig:I_V} presents current-voltage characteristics of a nanowire for different injection currents. The temperature and STM bias voltage were kept constant. The wire current $I_{wire}$ is successively ramped up and down while the wire voltage $V_{wire}$ is monitored. For no injection or low injection currents, i.e $I_t \leq 100$~pA, the current-voltage characteristic displays a clear superconducting behavior and a voltage jump occurs at the nanowire critical current $I_c$ when ramping up and at the retrapping current $I_r \leq I_c$ when ramping down. In the normal state, a resitive behavior is observed: $V_{wire}\approx R_N I_{wire}$ where $R_N$ is the nanowire normal state resistance. For high injected power, i.e for $I_t=2000$~pA, the current-voltage characteristic is almost purely resistive, suggesting that the power injected by the STM tip maintains the whole nanowire in the normal state. The actual injected power, $P \sim I_tV_b/2 =200$~pW is indeed close  to the minimum Joule heating required to keep the nanowire in the normal state:  $R_N I_r^2(I_t=0)=190$~pW.

\section{Supplementary Note 3: Scanning critical current microscopy}

\begin{figure}[h!]
	\centering
	\hskip -4.5mm 
	\resizebox{0.65\linewidth}{!}{\includegraphics{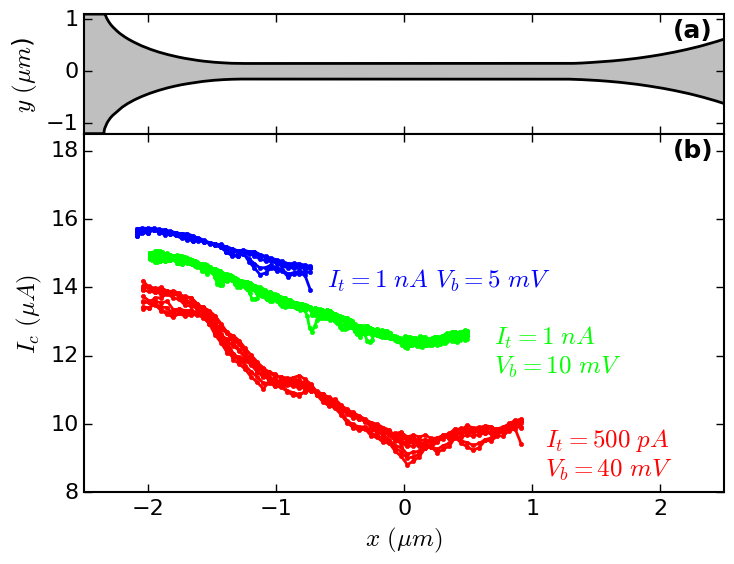}}
	
	\vskip -1.3mm
	\resizebox{0.65\linewidth}{!}{\includegraphics{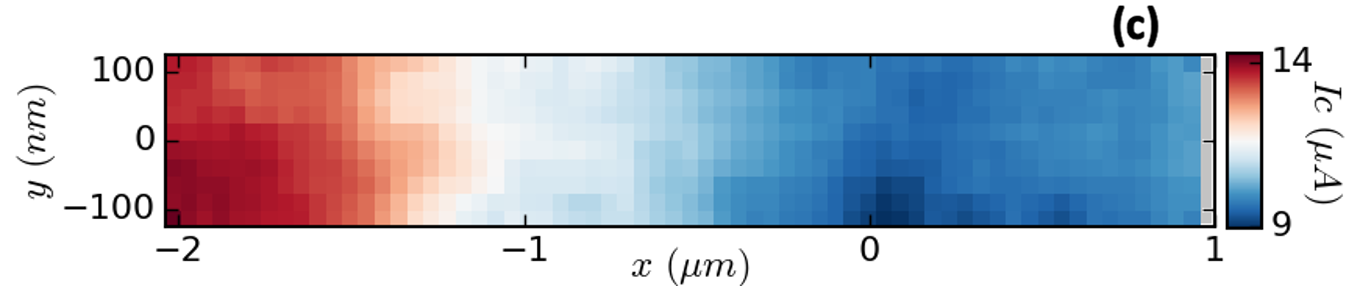}}
	\caption{\textbf{Influence of the injection position.} (a)~Nanowire shape. (b)~Critical current as a function of the STM tip position along the nanowire for different tunneling conditions at $T=180~mK$. (c)~Map of the critical current as a function of the injection position at a fixed tunneling setpoint: $I_t=500$~pA and $V_b=40$~mV. Without injection current, $I_c^0=18.5~\mu A$. Sample N06.}
	\label{fig:ic_x_N06}
\end{figure}

In contrast to lithographically defined tunnel junctions, scanning critical current microscopy allows to control the injection position of the quasiparticles with high spatial resolution. The STM tip scans the surface of the nanowire with a fixed tunneling setpoint, and stops at different positions to allow measuring the critical current. The scanning direction is parallel to the wire x-axis, see \autoref{fig:ic_x_N06}a.  Several lines are scanned from one side of the nanowire to the other. \autoref{fig:ic_x_N06}b displays the $x$-scans performed at various $y$-positions for several tunnelling conditions (blue, green and red curves). \autoref{fig:ic_x_N06}c shows a map of the critical current for the red tunneling setpoint, i.e  It = 500 pA and Vb = 40 mV.  The critical current depends mainly on the $x$-distance between the injection position and the leads, whereas the $y$-position has only a small influence. The scanning critical current microscopy results presented in Figure 3 of the main text are an average of these data along the $y$-direction.

\section{Supplementary Note 4: The phonon temperature}

In the analysis of the main text, the substrate plays the role of a reservoir at the bath temperature of the fridge $T_b$. We discuss here the validity of this hypothesis.

\begin{figure}[h!]
	\centering
	\resizebox{0.2\linewidth}{!}{\includegraphics{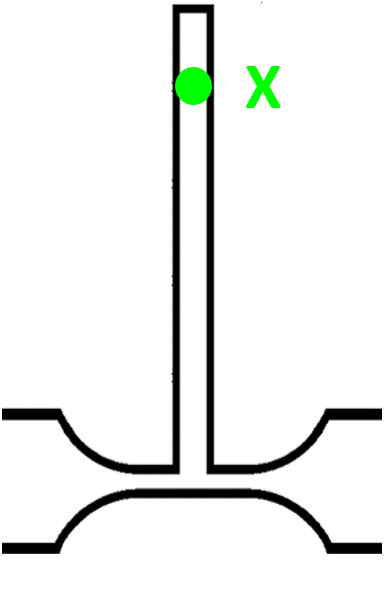}}
	\resizebox{0.6\linewidth}{!}{\includegraphics{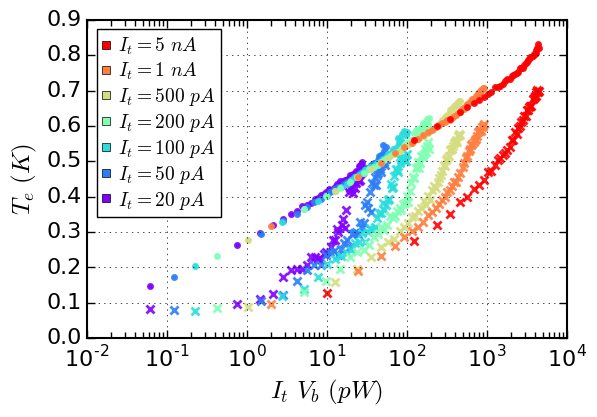}}
	\caption{\textbf{Effect of the injected power on the electronic temperature for injection of quasiparticles in the dead-end strip and in the ground plane.} (a)~Schematics of the sample with indicated injection positions. (b)~Electronic temperature at the crossing point between the nanowire and the dead-end strip as a function of $I_tV_b$ for injection in the dead-end strip (circles) and in the ground plane (crosses) at $T=70~$mK.}
	\label{fig:N16_substrate1}
\end{figure}

In this section, measurements are performed with a device constituted of a nanowire connected in its center to a dead-end strip as pictured on \autoref{fig:N16_substrate1}a. The film consists of a 3~nm layer of gold on top of a 3~nm layer of niobium so that its properties are comparable to the one of the samples presented in the main text: ${T_c=960}$~mK and $I_c^0=10.2$~$\mu$A.

\begin{figure}[h!]
	\centering
	\resizebox{0.6\linewidth}{!}{\includegraphics{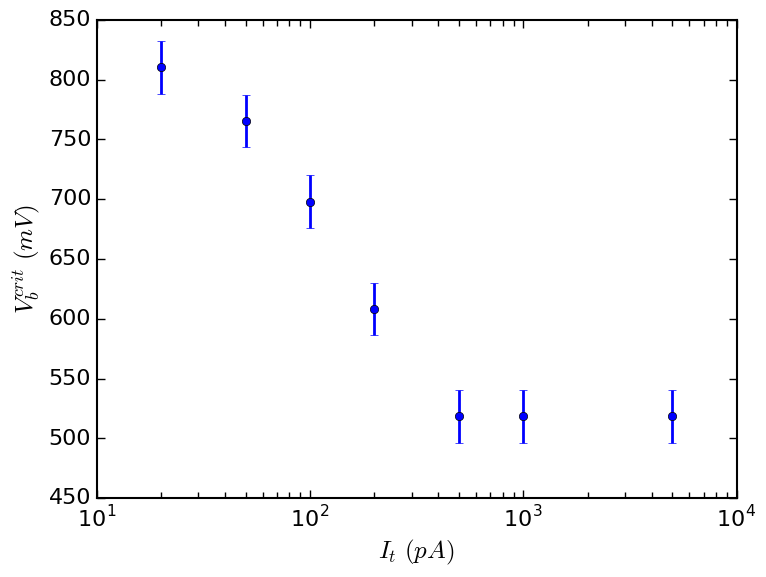}}
	\caption{\textbf{Phonon contribution to the heat flow.} Voltage threshold at different tunneling currents above which the assumption of a well thermalized  phonon bath breaks down.}
	\label{fig:phonon_threshold}
\end{figure}

We probed the limits of the reservoir approximation by comparing the effect of quasiparticle injection in the nanowire and in the ground plane surrounding it. In \autoref{fig:N16_substrate1}, panel~(a) pictures the injection positions. When quasiparticle injection is performed in the surrounding ground plane (colored crosses in \autoref{fig:N16_substrate1}b), which is electrically insulated from the nanowire, the electronic temperature of the nanowire is nevertheless affected. In this configuration, the entire heat flow between the ground plane and the nanowire is exclusively mediated through phonon conduction in the insulating substrate. Therefore, the latter necessarily holds a phonon temperature gradient and we have to drop the assumption $T_{ph} = T_b$ everywhere in the sample. Furthermore, as shown in \autoref{fig:N16_substrate1}b, the critical current is not unequivocally determined by the injected power since $T_e$ can take different values for a given $I_tV_b$ depending on the tunneling current $I_t$. In contrast, when injection is performed in the dead-end strip (colored disks), one recovers the thermal behavior presented in the main text, where the electronic temperature depends linearly on the logarithm of $I_tV_b$ up to very large voltages where it starts to deviate. Strikingly, this deviation occurs approximately when curves of crosses and dots intersect, \textit{i.e.} when the electronic temperature elevation in the nanowire due to electron conduction and to phonon conduction in the substrate are comparable. For each tunneling current, the voltage threshold $V_b^{crit}$ above which deviations to the thermal behavior are observed is shown in \autoref{fig:phonon_threshold}. Below this threshold, the main thermal conduction between the injection point and the nanowire is insured by the electrons and the phonon bath can be considered as a reservoir. In the main text, $V_b$ never exceeds 400 mV which is lower than $V_b^{crit}$ for any tunnel current. Our assumption that the substrate acts as a thermal reservoir for phonons ($T_{ph}=T_b$) is therefore valid.

\newpage
\section{Supplementary Note 5: Electronic thermal conductivity in the superconducting state within Usadel framework}

The development of the quasi-classical theory of superconductivity starting from non equilibrium Green functions in the Keldysh formalism is presented in~\cite{chandrasekhar_Introduction_2004}. The diffusive limit leads to the Usadel equations which merge in an unified formalism for both equilibrium and out of equilibrium properties. The physical quantities of interest can be computed from retarded and advanced Green functions $R$ and $A$ (which are $2\times2$ matrices depending on position and energy) and filling factor $h$. However, the parameterized expression of the thermal conductivity is not available in~\cite{chandrasekhar_Introduction_2004} and is therefore derived here.\\

The filling factor can be decomposed in an odd $h_{od}$ and an even $h_{ev}$ functions of the energy, and is related to the distribution function of electrons $f$ through $h=1-2f$. The retarded and advanced Green functions $R$ and $A$ must obey some normalization conditions and can be parameterized as follows~\cite{sohn_Mesoscopic_2013,chandrasekhar_Introduction_2004}:

\begin{equation}
	R=
	\begin{pmatrix}
		\cos\theta&\sin\theta e^{i\phi}\\
		\sin\theta e^{-i\phi}&-\cos\theta
	\end{pmatrix}
\end{equation}

\begin{equation}
	A=-\tau_3R^\dagger\tau_3
\end{equation}
where $\dagger$ denotes the Hermitian conjugate, and the $\tau$ matrices are equal to the identity and Pauli matrices:
\begin{equation}
	\begin{array}{cc}
		\tau_0=
		\begin{pmatrix}
			1&0\\
			0&1
		\end{pmatrix},&
		\tau_1=
		\begin{pmatrix}
			0&1\\
			1&0
		\end{pmatrix},\\
		\tau_2=
		\begin{pmatrix}
			0&-i\\
			i&0
		\end{pmatrix},&
		\tau_3=
		\begin{pmatrix}
			1&0\\
			0&-1
		\end{pmatrix}
	\end{array}
\end{equation}
so that:
\begin{equation}
	A=
	\begin{pmatrix}
		-\cos\theta^*&\sin\theta^* e^{i\phi^*}\\
		\sin\theta^* e^{-i\phi^*}&\cos\theta^*
	\end{pmatrix}
\end{equation}\\

Defining the quantities 
\begin{equation}
	\begin{aligned}
		\overrightarrow{Q}&=\frac{1}{4}\textnormal{Tr}\left\lbrace \tau_3\left(R\overrightarrow{\nabla}R-A\overrightarrow{\nabla}{A}\right) \right\rbrace\\
	\end{aligned}
\end{equation}

and
\begin{equation}
	M_{ij}=\frac{1}{4}\textnormal{Tr}\left\lbrace \delta_{ij}\tau_0-R\tau_iA\tau_j \right\rbrace
\end{equation}
where $\delta_{ij}$ is the Kronecker delta, the thermal current can be written as:
\begin{equation}
	\overrightarrow{J_{th}}=\frac{\sigma_N}{2e^2}\int_{-\infty}^{\infty}d\epsilon~\epsilon\left( M_{00}\overrightarrow{\nabla}h_{od}+\overrightarrow{Q}h_{ev}+M_{30}\overrightarrow{\nabla}h_{ev} \right)
\end{equation}
with $\sigma_N$ the normal state conductivity.

When the system is at local thermal equilibrium, $h_{od}(x,\epsilon)=\tanh(\epsilon/2k_BT(x))$ and $h_{ev}(x,\epsilon)=0$.
Then:
\begin{equation}
	\begin{aligned}
		\overrightarrow{\nabla}h_{od}=&\frac{dh_{od}}{dT}\overrightarrow{\nabla}T\\
		=&\frac{-\epsilon}{2k_BT^2}\left(1-\tanh^2\left(\frac{\epsilon}{2k_BT}\right)\right)\overrightarrow{\nabla}T
	\end{aligned}
\end{equation}

For a bulk superconductor for which $\phi$ is real, $M_{03}=M_{30}=0$, $M_{33}=\cosh^2(\textnormal{Im}[\theta])$ and $M_{00}=\cos^2(\textnormal{Re}[\theta])$.\\

Finally, in the quasi-equilibrium limit one obtains $\overrightarrow{J_{th}}=-k_e\overrightarrow{\nabla}T$ with the electronic thermal conductivity equals to :

\begin{equation}
	k_e=\frac{\sigma_N}{2e^2}\int_{-\infty}^{\infty}d\epsilon~\frac{\epsilon^2}{2k_BT^2}
	\left(1-\tanh{}^2\left(\frac{\epsilon}{2k_BT}\right)\right)\times\cos^2(\textnormal{Re}[\theta])
\end{equation}

which value at critical temperature ($\theta=0$) corresponds to Wiedemann-Franz law:
\begin{equation}
	\frac{k_e}{\sigma_NT}=\frac{\pi^2}{3}\left(\frac{k_B}{e}\right)^2
	\label{eq:Wiedemann_Franz}
\end{equation}

From the above equation, the electronic thermal conductivity can be computed for any value of the temperature and of the supercurrent flow since the pairing angle $\theta$ is given by Usadel equations with an homogeneous current distribution~\cite{usadel_Generalized_1970,belzig_Quasiclassical_1999}:
\begin{subequations}
	\begin{align}
		&\epsilon+i\Gamma_{in}+i\gamma\cos\theta=i\Delta\frac{cos\theta}{\sin\theta}\label{eq:usadel}\\
		&\Delta=N_0V_{eff}\int_{0}^{\hbar\omega_D}d\epsilon\tanh\left(\frac{\epsilon}{2k_BT}\right)\textnormal{Im}[\sin\theta]
		\label{eq:delta}
	\end{align}
\end{subequations}
with $N_0$ the density of states at Fermi level, $V_{eff}$ the BCS pairing potential, $\omega_D$ the Debye frequency, $\Gamma_{in}$ the inelastic scattering term and $\gamma$ the depairing energy given by:
\begin{subequations}
	\begin{align}
		\frac{J_s}{J_\gamma}&=\sqrt{\frac{\gamma}{\Delta_0}}\frac{U_s}{\Delta_0}\label{eq:igamma}\\
		U_s&=\int_{0}^{\infty}d\epsilon \tanh\left(\frac{\epsilon}{2k_BT}\right)\textnormal{Im}[\sin^2\theta]
	\end{align}
\end{subequations}
where $\Delta_0$ is the zero temperature superconducting gap (which is equal to the order parameter in the absence of depairing), $J_s$ is the supercurrent density and ${J_\gamma=\sqrt{2}\Delta_0^{3/2}\sqrt{N_0\sigma_N/\hbar}}$ is a material dependent parameter directly related to the zero temperature critical current.

The homogeneous current distribution approximation is valid when the London penetration depth is large compared to the transverse dimensions of the nanowire.
In a thin film, the effective London penetration depth is~\cite{klein_Effective_1990} $\lambda_L^2/d$ with $\lambda_L=\sqrt{\hbar/(\pi\mu_0\sigma_N\Delta_0)}$ and $d$ the thickness of the film.
In our samples, this leads to an effective penetration depth of about 1 to 10~$\mu$m, whereas our nanowires are 300~nm wide and about 10~nm thick, so that the hypothesis of homogeneous current distribution is valid.

\section{Supplementary Note 6: Simulation of the injected quasiparticles dynamics}

In order to have a quantitative analysis of the quasiparticles dynamics we performed numerical simulations according to the hotspot model of Semenov and co-workers~\cite{semenov_spectral_2005}. The critical current of the nanowire is given by :
\begin{eqnarray}
	I_c(t) = I_c^{stat} \, (1 - \frac{(1-e^{\frac{-t}{\tau_{rel} } })\, e^{\frac{-t}{\tau_{rec}}}}{N_0 \, \Delta_0  \, w \,  d \, \sqrt{\pi \, D \, t}})
	\label{eq:Ic_vs_deltaN}
\end{eqnarray}

\begin{figure}[!h]
	\resizebox{0.7\linewidth}{!}{\includegraphics{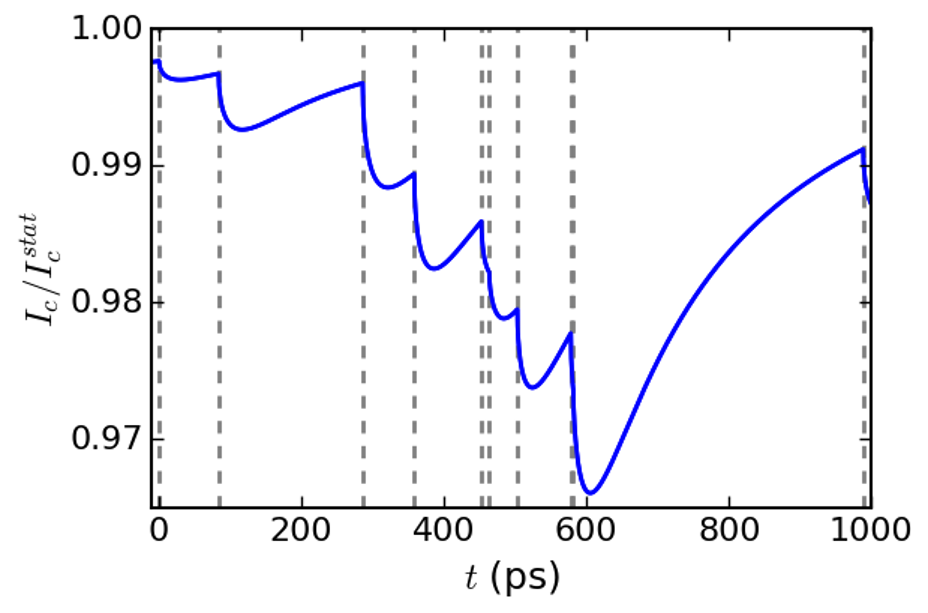}}
	\caption{\textbf{Time evolution $I_c(t)$.}  Example of a burst of injected quasiparticles, each of them occurring at a time pointed out by a vertical dashed line.}
	\label{Ic(t)}
\end{figure}

The total number of quasiparticles in excess results from the addition of several tunneling events which occur at random times obeying a Poisson distribution where on average one electron is tunneling every $\tau_{inj}=\frac{e}{I_t}$. The energy of the tunnelling electron is also random and uniformly distributed between $0$ and $eV_b$. For each series of $N = 10^7$ tunnelling events, we found that the minimum critical current corresponds to a burst of seven or eight tunneling events similar to what is shown in \autoref{Ic(t)}. The time between each of them is short enough so that all the generated quasiparticles do not have enough time to diffuse out of the critical volume. This leads to an important accumulation of quasiparticles and therefore a minimum of the critical current.

\begin{figure}[!h]
	\resizebox{0.7\linewidth}{!}{\includegraphics{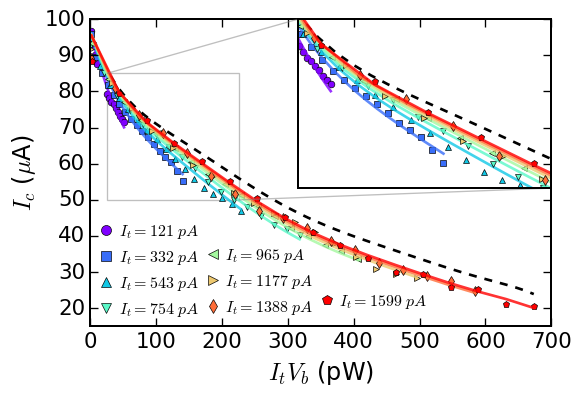}}
	\caption{\textbf{Relaxation dynamics of the injected quasiparticles for a different recombination time.}  Points are the same data as in Figure 5 of the main text and solid lines are numerical fits with $\tau_{rel} = 40$~ps and $\tau_{rec} = 250$~ps. The dashed line is the stationary critical current $I_c^{stat}$. The inset is a zoom at low power.}
	\label{recombination}
\end{figure}

The recombination time scale $\tau_{rec}$ acts as a time cutoff, avoiding quasiparticles accumulation in the nanowire for long computation times. The results shown in the main text have been computed with $\tau_{rec} = 500$~ps which is a typical recombination time~\cite{kaplan_quasiparticle_1976}. We checked that the fitting value of $\tau_{rel}$  is only slightly dependent on the choice of $\tau_{rec}$. This is illustrated in \autoref{recombination} which shows the critical current as a function of $I_t V_b$ computed with $\tau_{rel}= 40$~ps and $\tau_{rec} = 250$~ps. The results are very similar to those shown in the main text with $\tau_{rel}=40$~ps  and $\tau_{rec}=500$~ps.

\begin{figure}[!h]
	\resizebox{0.6\linewidth}{!}{\includegraphics{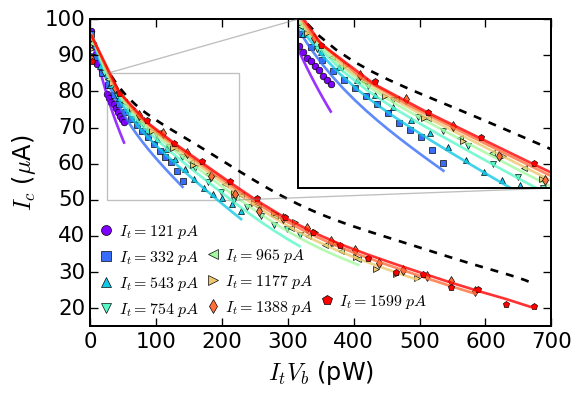}}
	\resizebox{0.6\linewidth}{!}{\includegraphics{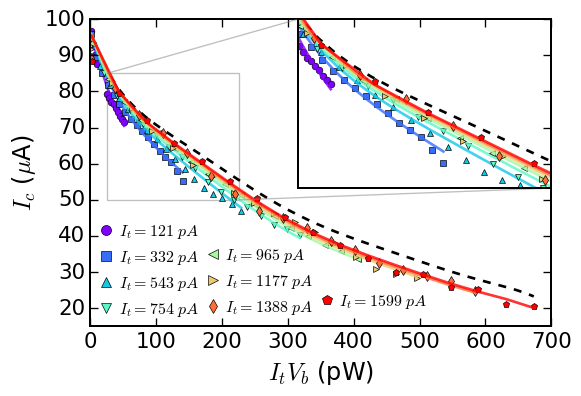}}
	\caption{\textbf{Relaxation dynamics of the injected quasiparticles for different relaxation times.}  Points are the same data as in Figure 5 of the main text and solid lines are numerical fits with $\tau_{rel} = 20$~ps (top) and $\tau_{rel} = 60$~ps (down). The dashed line is the stationary critical current $I_c^{stat}$. The inset is a zoom at low power.}
	\label{relaxation}
\end{figure}

In order to estimate the accuracy of our fit in the determination of $\tau_{rel}$, we varied its  value. \autoref{relaxation} displays two simulations with $\tau_{rel} = 20$~ps and $\tau_{rel} = 60$~ps. For $\tau_{rel} = 20$~ps the computed critical current is much lower than the experimental values at low tunnelling current, whereas for $\tau_{rel}=60$~ps the discrepancy between experiment and computation is more pronounced at higher $I_t$. We therefore concluded that the best trade off is about $\tau_{rel} = 40  \pm 10$~ps. This rather large uncertainty may come from the assumption that $\tau_{rel}$ does not depend on the energy of the injected quasiparticles in the theoretical model and all the data are therefore fitted with the same relaxation time.

\end{document}